# Lateral modulation of magnetic anisotropy in tricolor 3*d*-5*d* oxide superlattices

*Zengxing Lu,* [†,§] *Jingwu Liu,* [†,§,ξ] *Lijie Wen,* [†,§,ε] *Jiatai Feng,* [†,§] *Shuai Kong,* [†,§] *Xuan Zhen,* [†,§,δ] *Sheng Li,* [†,§] *Peiheng Jiang,* [†,§] *Zhicheng Zhong,* [†,§,‡] *Junfa Zhu,* [ξ] *Xianfeng Hao,* [*,ε] *Zhiming Wang,* [*,†,§,‡] *and Run-Wei Li* [†,§,‡]

[†]*CAS Key Laboratory of Magnetic Materials and Devices, Ningbo Institute of Materials Technology and Engineering, Chinese Academy of Sciences, Ningbo 315201, China*

[§]*Zhejiang Province Key Laboratory of Magnetic Materials and Application Technology, Ningbo Institute of Materials Technology and Engineering, Chinese Academy of Sciences, Ningbo 315201, China*

[ξ]*Nano Science and Technology Institute, University of Science and Technology of China, Hefei 230026, China*

[ε]*Key Laboratory of Applied Chemistry, College of Environmental and Chemical Engineering, Yanshan University, Qinhuangdao 066004, China*

[δ]*University of Nottingham, Ningbo 315201, China*

[‡]*Center of Materials Science and Optoelectronics Engineering, University of Chinese Academy of Sciences, Beijing 100049, China*

**ABSTRACT:** Manipulating magnetic anisotropy (MA) purposefully in transition metal oxides (TMOs) enables the development of oxide-based spintronic devices with practical applications. Here, we report a pathway to reversibly switch the lateral magnetic easy-axis via interfacial oxygen octahedral coupling (OOC) effects in 3*d*-5*d* tricolor superlattices, i.e. $[SrIrO_3,mRTiO_3,SrIrO_3,2La_{0.67}Sr_{0.33}MnO_3]_{10}$ ($RTiO_3$: $SrTiO_3$ and $CaTiO_3$). In the heterostructures, the anisotropy energy (MAE) is enhanced over one magnitude to ~$10^6$ erg/$cm^3$ compared to $La_{0.67}Sr_{0.33}MnO_3$ films. Moreover, the magnetic easy-axis is reversibly reoriented between (100)- and (110)-directions by changing the $RTiO_3$. Using first-principles density functional theory calculations, we find that the $SrIrO_3$ owns a large single-ion anisotropy due to its strong spin-orbit interaction. This anisotropy can be reversibly controlled by the OOC, then reorient the easy-axis of the superlattices. Additionally, it enlarges the MAE of the films via the cooperation with a robust orbital hybridization between the Ir and Mn atoms. Our results indicate that the tricolor superlattices consisting of 3*d* and 5*d* oxides provide a powerful platform to study the MA and develop oxide-based spintronic devices.

**Keywords:** iridates, magnetic anisotropy, tricolor superlattice, spin-orbit coupling, oxygen octahedral coupling

## INTRODUCTION

Magnetic anisotropy (MA) plays a key role in enriching the spectrum of physical responses and enabling high-performance spin devices.[1, 2] It is important to develop effective strategies to manipulate the MA in magnetic systems, including changing magnetic easy-axis and MA energy (MAE). Transition metal oxides (TMOs) heterostructures and superlattices provide an intriguing playground for manipulation of electronic and magnetic properties.[3-6] Advances in thin-film fabricating techniques enable atomically precise fabrication of artificial heterostructures and superlattices comprising dissimilar oxides with strong magnetic and spin-orbital interactions, which are fundamentally important for achieving tunable and strong MA.[7-13] For instance, superlattices consisting of two different components, 3*d* oxide $La_{1-x}Sr_xMnO_3$ ( $0 \leq x \leq 1$) and 5*d* oxide $SrIrO_3$ (SIO), exhibit strong magnetic anisotropy with an enhanced magnitude from $10^5$ to $10^6$ erg/cm$^3$.[11, 12] To further modulate the magnetic easy-axis, two main approaches are developed through controlling the cation doping level in 3*d* $La_{1-x}Sr_xMnO_3$[12] and the dimensionality of 5*d* SIO[10, 11, 13]. However, both approaches significantly weaken the magnetic interactions in $La_{1-x}Sr_xMnO_3$, thereby reducing the Curie temperature and magnetism dramatically. Moreover, changing the dimensionality in SIO will suppress the emergent Ir magnetization and the associated large single-ion anisotropy.[13-15] Therefore, it is challenging to achieve both strong MAE and tunable easy-axis only by adjusting the two components in these binary superlattices. It is of great importance to develop an efficient method to modulate the easy-axis while enhancing the MAE.

It has been demonstrated that tricolor superlattices composed of three different component layers provide a powerful platform, leading to a rich spectrum of novel phenomenon.[5, 6, 16] Prominent examples include induced and/or enhanced ferroelectricity, magnetoelectric multiferroicity, novel spin and orbital ordered states in 3*d* titanates[16-18], manganites[19, 20], and nickelates[21, 22]. In tricolor superlattices, multiple heterointerfaces exist between dissimilar oxides, enabling more flexible engineering the structural, electronic and magnetic coupling. Previous studies have revealed the important role of structural coupling for tuning MA in $La_{1-x}Sr_xMnO_3$ and $SrRuO_3$, especially through oxygen octahedral coupling (OOC).[23-26] Intriguingly, recent studies have revealed that the OOC shows impact on the MA of 5*d* SIO too.[27] Considering that the $La_{1-x}Sr_xMnO_3$ possessing large magnetic moment but small MAE, while the SIO possessing small magnetic moment but large single-ion anisotropy[13], it is of great importance to investigate the cooperative effect of the MAs of individual 3*d* and 5*d* components modified by the OOC in $La_{1-x}Sr_xMnO_3$/SIO heterostructures. This effect will enrich the control strategies in the MA, however, there is few studies exploring the cooperative effect. Tricolor superlattices consisting of 3*d* and 5*d* oxides provide high flexibility to control structural, electronic and magnetic coupling, thereby offering an opportunity to investigate the cooperative effect on the MAs in the TMOs.

In this work, we have fabricated $La_{0.67}Sr_{0.33}MnO_3$-SIO-$RTiO_3$ ($RTiO_3$: $SrTiO_3$ and $CaTiO_3$) tricolor superlattices and investigated the cooperative effect on in-plane MA via OOC engineering. Firstly, in these superstructures, the MAE is enhanced over one order of magnitude to ~$10^6$ erg/$cm^3$ compared in pure LSMO. Secondly, the magnetic easy-axis is reversibly reoriented between (100) and (110) directions simultaneously. Combining the structural analysis and first-principle calculations, we find that the $MnO_6$ and $IrO_6$ octahedral rotation modes can be tuned via the OOC by altering $SrTiO_3$ (STO) and $CaTiO_3$ (CTO), and there is a large single-ion anisotropy in the SIO. The anisotropy of the SIO rather than of the LSMO can be adjusted by the tunable structure, then reorients the easy-axis of the superstructure reversibly. At the meantime, a strongly orbital hybridization exists between the Ir and Mn atoms at the SIO/LSMO, and enhances the MAE via the cooperation with the single-ion anisotropy. Our results indicate that constructing tricolor superlattice with 3*d* and 5*d* oxides provides a flexible platform to study the MA on which one can continuously control the OOC through properly structural design, then consecutively tune the easy-axis in one system with robust MA, which is important to control and induce noncollinear magnetic structures, such as spiral texture and skyrmion, and further develop spin devices.

## MATERIALS AND METHODS

*Superlattice preparation.* A serious of $SrIrO_3$-$La_{0.67}Sr_{0.33}MnO_3$ (SIO-LSMO) superlattices (SLs) were designed with a configuration of $[SIO,mRTO,SIO,2LSMO]_{10}$ (RTO: $SrTiO_3$ and $CaTiO_3$, m=0, 1, 2, 3). Figure 1a shows schematic diagram of one unit building block of the superlattices (m=3). The SLs were epitaxially grown on the single crystal (001) $SrTiO_3$ (STO) substrates by pulsed laser deposition (PLD). The substrates were etched by buffered hydrofluoric acid and annealed in oxygen atmosphere to get an atomically flat and $TiO_2$-terminated surface. A single-crystal STO and SIO, LSMO, $CaTiO_3$ (CTO) ceramic targets were used to deposited the films.

The films were prepared using a KrF (λ=248 nm) excimer laser with a laser energy density of 1.3 J/cm$^2$ and a repetition rate of 2 Hz under an oxygen pressure of 0.1 mbar at 680 ℃. Then the SLs were cooled to room temperature with a cooling rate of 5 ℃/min under an oxygen pressure of 1 mbar. During depositing, the growth date was monitored by reflection high-energy electron diffraction (RHHED) oscillation.

*Structure and magnetism characterizations.* The surface quality and oxygen octahedral rotation (OOR) were analyzed by the RHEED patterns captured under high vacuum after cooling to room temperature. For the structural study via the RHHED, all of the samples were terminated with LSMO and/or SIO layers, and the patterns were shot at the LSMO and/or SIO surfaces. The information on the phase purity, crystal structure and epitaxial quality were examined by X-ray diffraction (XRD) scan (D8 Discover, Bruker). The magnetic anisotropy (MA) was studied with the magnetization loops ($\mu_0H$-$M$ loops) and temperature-dependent magnetization curves ($M$-$T$ curves) along different crystalline orientations of (001), (100), (010) an (110). The loops and curves were measured using a superconducting quantum interference device (SQUID, Quantum Designer Inc.). Anisotropy magnetoresistance (AMR) measurements were performed with physical property measurement system (PPMS, Quantum Designer Inc.). The transport measurements were conducted in a van der Pauw method with Ti contacts capped with Pt electrodes under planar magnetic field. The relative angle between the field and the (100)-crystalline orientation was controlled by rotating the sample holder.

*First-principles calculation method:* First principles calculations were performed within the Vienna ab initio simulation package (VASP) code[28-30]. In all the calculations, the exchange-correlation potential was treated with the generalized gradient approximation (GGA) with the Perdew-Burke-Ernzerhof (PBE) functional[31]. The strong correlation effects for the Mn 3d and Ir 5d electrons are taken into account by means of the GGA+U approach and the on-site effective Coulomb interaction parameter $U_{eff}$ is taken as 4.0 eV and 2.0 eV, respectively[32]. The cutoff energy was 400 eV and 3×3×1 k-mesh was used in all simulations. Two superlattice models as illustrated in Fig. S6, i.e., a “real” distorted interface and an “ideal” undistorted one, consists of alternative symmetrical 3 layers LSMO and SIO components. Note that, the prominent distinction of these two models is the $IrO_6$ and $MnO_6$ octahedra are allowed to rotation with the $a^+b^-c^-$ pattern in the former case, compared to the latter one. The planar lattice parameters are fixed to the lattice constant of $SrTiO_3$ (001) surface within $\sqrt{2}\times\sqrt{2}$ framework (~7.81 Å), while the vertical axis has been optimized until the force on each atom is smaller than 0.01 eV/Å. Furthermore, the magnetic anisotropy energy is calculated by using the force theorem[2].

## RESULTS AND DISCUSSION

The crystal structures are characterized by the $\theta$-$2\theta$ XRD. Figure 1b-d show the XRD patterns of $[2SIO,2LSMO]_{10}$ (SL), $[SIO,3STO,SIO,2LSMO]_{10}$ (SL-3S) and $[SIO,3CTO,SIO,2LSMO]_{10}$ (SL-3C) films, respectively. In addition, X-ray reciprocal space maps (RSMs) are taken near the (103) reflection peak of the STO substrate for the SL-3S (Figure 1e) and SL-3C (Figure 1f) heterostructures. We note that several superstructure peaks of the SL-3S are weak (Figure 1c), leading that the $SL_{-1}$-3S peak in RSM is undetectable (Figure 1e). However, other XRD peaks are pronounced comparable to that of the SL-3C, indicating the sharp interfaces and strict periodicity in above superlattices. As shown in Figure 1e,f, the vertical alignment of the peaks of the superlattices and substrates suggest that the films are fully strained by the substrates. It should be noted that the CTO is an orthorhombic and its pseudo-cubic lattice constant is 3.847 Å smaller than the cubic STO value (3.905 Å).[33, 34] So the SL-3C has a thinner unit of building block than the SL-3S. This is confirmed by in the XRD and RMS characterizations in which the SL-3C has larger values (labelled as $SL_{-1}$-3C and $SL_{-2}$-3C) of $2\theta$ and $l$ in Figure 1d,f. The above results demonstrate that the superlattices have been epitaxially grown with high quality and well-defined periodicity.

To study the MA of the films, the measurement of magnetization loops ($\mu_0H$-$M$ loops) is performed along different axes (i.e. (100), (010), (110) and (001)) of LSMO and SL films. Comparing those loops, we find that the LSMO is soft ferromagnetic with small coercive field about $10^{-3}$ T scale, and has an in-plane easy-axis along (110) direction, as shown in Figure S1b and Figure 2a. When inserting 2 u.c. SIO between each 2 u.c. LSMO, it becomes harder with a coercive field about $10^{-1}$ T scale (Figure S1c). We further confirm that the easy-axes lie in the film planes, seen from Figure S1. Note that the easy-axis of LSMO totally lies in plane with negligible out-of-plane component. While the SL shows a partial spontaneous magnetization along (001) axis, i.e. toward out-of-plane, which may be owed to the enlarged magnetocrystalline anisotropy caused by the strong SOC of SIO and partially cancel out the shape anisotropy of LSMO.[11-13] Even so, the adjusting effect of SIO works on the in-plane MA more obviously. As shown in Figure 2b, its easy-axis is reoriented to (100) and equivalent (010) directions, the same as the observation in Ref. 13. These indicate that the SIO has a great influence on the MA of LSMO.

Intriguingly, the switching of the in-plane easy-axis can be reversed. When further inserting 1 u.c. STO in the SL to develop a new superstructure, $[SIO,STO,SIO,2LSMO]_{10}$ (SL-S), the magnetic easy-axis is switched back to (110) (Figure S2b). This modulation becomes more prominent after inserting 2 and 3 u.c. STO in $[SIO,2STO,SIO,2LSMO]_{10}$ (SL-2S) and SL-3S, as shown in Figure S2c and Figure 2c, respectively. However, when replacing the STO with another titanate, CTO, the axis is reoriented to (100) in SL-3C (Figure 2d) as same as in the SL film. The temperature-dependent magnetizations (*M-T*) show same results (Figure S3). This reversible tunable effect is further validated by the test of anisotropic magnetoresistance (AMR), as exhibited in Figure S4. In Figure 2e, the AMR curves of LSMO, SL, 3STO-SL and 3CTO-SL are plotted with polar form. Firstly, we find that all of the plots show quasi-fourfold rotation, indicating the same symmetry of the MA. Secondly, there is an approximate 45° shift of AMR between LSMO and SL. When inserting STO and CTO layers, the reversible 45° rotation is also observed too. These are coincident with the MA evolution shown in Figure 2a-d. Figure 2f summarizes the impact of the SIO and RTO layers on the evolution of LSMO-MA. Firstly, the MAE (defined as $E_{(100)}$ - $E_{(110)}$) is significantly enhanced more than one order of magnitude to ~$10^6$ erg/$cm^3$ by building the bicolor and/or tricolor LSMO-SIO superlattices. Secondly, the MAE of the STO-based superlattices increases with increasing the thickness of STO. Thirdly, the easy-axis can be modulated not only by the SIO but also by the inserting RTOs (STO and CTO which have different crystal structures)[33-35]. It indicates that there are other possible causes affecting the MA besides the strong interfacial magnetic coupling at the SIO/LSMO interface proposed in Ref. 13.

To reveal the origin of the reversible tuning of MA, we first examine the structural evolution upon inserting varying RTO layers in the SIO/LSMO superlattices. *In-situ* RHEED measurements were performed on LSMO-terminated superlattices to investigate the evolution of $MnO_6$ octahedral rotations upon inserting the $TiO_6$ octahedron. After depositing the SL-3S, SL and SL-3C (LSMO-terminated), the patterns are captured along (110) and (100) directions, as shown in Figure 3a,d,e, respectively. In the (100)- and (110)-pattern images of SL-3S (Figure 3a), the streaks marked with white arrows exhibit a 1×1 periodicity in reciprocal space (Figure 3b) at the film surface. Similar result is observed in the SL and SL-3C (Figure 3d,e). However, several gracile streaks (marked with red arrows) exist between the 1×1 spots in the (110) image of the SL and SL-3C films while not exist in the SL-3S. Viewed along (110) direction, the observed gracile streaks correspond to a $\sqrt{2} \times \sqrt{2}$ surface periodicity (Figure 3f) which is the manifestation of lattice distortion due to the oxygen octahedral rotation (OOR)[36], as illustrated in Figure 3g. At the meantime, as shown in Figure S5, we capture the patterns of the 3STO/SIO and 3CTO/SIO bilayers (SIO-terminated), and find that the $IrO_6$ can be adjusted by the RTOs too. Therefore, these results reveal that the controllable OOR in the tricolor superlattices can be achieved by inserting different RTO layers, which in turn give rise to a reversible reorientation of MA.

To understand the correlation between reversible tuning of MA and the structural evolution, we performed density functional theory (DFT) calculations of the MAE. To simplify the structural model while still capture the main structural characteristics of SL-3C and SL-3S, we construct 3LSMO/3SIO supercells with and without OOR shown in Figure S7 (see details in Section V of Supporting Information), denoted as distorted and undistorted, respectively. According to Figure 4b, firstly, our calculation indicates that the in-plane MAE is very large, and up to 0.53 and 0.77 meV/u.c. ($\sim 1.4\times10^7$ and $\sim 2.1\times10^7$ erg/cm$^3$) in cells with and without distortion, respectively, which is two orders of magnitude higher than that in LSMO with an order of $10^{-3}$ meV/u.c.[23]. This is consistent with literatures that the SIO layer with strong SOC can significantly enlarge the magnetic anisotropy energy.[11-13] More importantly, the easy-axis can be switched in the plane when altering the supercell structure. For the distorted case, it shows a MA with the easy-axis along (100) direction. While for the undistorted one, the easy-axis is oriented along (110). These results are in good agreement with our experimental observations, confirming that the tunable OOR plays a dominant role in determining the reversible MA in tricolor SL-3C and SL-3S superlattices.

To gain further insight into the origin of reversible MA, we have performed layer-, element- and orbital-projected MAE calculations of the 3LSMO/3SIO supercell based on a second-order perturbation theory (see details in METHODS). We define the MAE along the (100)-axis as positive and (110)-axis as negative. According to Figure S8 and Figure S9, the layer-projected MAE calculations of the SIO show that the interfacial layer exhibits larger MAE value than that in the internal layer, while the element-projected MAE calculations show the value of Ir-MAE is two magnitudes larger than that of the Mn-MAE. This is consistent with the scenario that the SIO possesses a large single ion anisotropy due to the interplay between strong SOC and crystal field splitting[13]. Moreover, we note that a strong coupling between Ir and Mn exists at the interface from partial density of states (PDOS) and differential charge density calculations. As shown in Figure 4e, the Ir-DOS at the Fermi level is derived from the Ir $t_{2g}$ orbits, while in Figure 4f, the Mn $e_g$ orbital states locate around the Fermi level, indicating that there is an orbital hybridization via possibly forming an interfacial molecular orbit between Ir and Mn atoms[14, 37]. The hybridization can be directly visualized from the calculated differential charge density distribution too (see Figure S10). Therefore, the combination of strongly interfacial hybridization and large single ion anisotropy will enable a pronounced response to the external stimulus and provide a strong MAE.

More importantly, according to the orbital-resolved Ir-MAE calculations, we find that the large single ion anisotropy of Ir atoms can be reversibly tuned by changing the crystal structure. Figure 4c,d show the MAE contributed by the interfacial SIO with the value of 4.633 and -2.083 meV for the distorted and undistorted structures, respectively. The positive sign of Ir-MAE appears in the distorted structure while it turns into negative in the undistorted structure, meaning that the easy-axis of the SIO with single-ion anisotropy can be switched from (100) to (110) by eliminating the structural distortion. The largest MAE change comes from the ($d_{xy}$, $d_{xz}$) matrix element, where the positive value in the structure with distortion reverses to negative value without distortion, and resulting in a sign change of Ir-MAE. In sharp contrast, the sign of Mn-MAE keeps negative regardless of the structural distortion (see Figure S9c,d). These distinct differences demonstrate that the tunable MA is originated from the change of interfacial Ir-MAE. So in our heterosystems, their MA can be modulated by reorienting the easy-axis between (100) and (110) of anisotropy in the SIO rather than in the LSMO via the interfacial OOC which controls the structure, leading to the reversible switching of the easy-axis in the films.

## CONCLUSION

In conclusion, we present the importance of the tricolor superlattices containing 5*d* heavy metal oxides with strong SOC to enhance and tune the MA of TMOs via the interfacial OOC. In these heterostructures, i.e. LSMO-SIO-RTO superlattices, there is a large single-ion anisotropy induced by the robust SOC in the interfacial SIO and a strongly orbital hybridization between the Ir and Mn atoms at the SIO/LSMO. The former one is sensitive to the structural transformation and can be modulated by altering the RTO (STO and CTO), then reversibly switches the MA of the superlattice films. In the SL-3C, which has a distorted structure with rotating octahedron, the SIO shows an anisotropy with a preferred (100)-orientation, leading to a (100)-oriented MA in the film. On the contrary, in the SL-3S, the anisotropy becomes (110)-oriented accompanying the suppressing effect of octahedral rotation caused by the STO, then switches the easy-axis of the film to (110) direction. At the meantime, because of the cooperative effect of the hybridization and the single-ion anisotropy, the MA of the films is enhanced greatly too. This study indicates that the tricolor superlattices and OOC effects are of great significance for the MA in the TMOs. And it provides a feasible route to continuously manipulate the easy-axis in one system with robust MA through properly configural design, then control and induce noncollinear magnetic structures and further design emerging spin devices.

ASSOCIATED CONTENT

**Supporting information:** (1) fabrication and magnetic anisotropy (MA) of LSMO and $[2SIO,2LSMO]_{10}$ superlattice films; (2) structure and MA of $[SIO,STO,SIO,2LSMO]_{10}$ (SL-S) and $[SIO,2STO,SIO,2LSMO]_{10}$ (SL-2S) superlattice films; (3) in-plane magnetic anisotropy of the LSMO and superlattice films characterized by the measurements of temperature-dependent magnetization and anisotropic magnetoresistance; (4) in-plane structural transition of the SIO modulated by STO and CTO; (5) magnetic ground states for the distorted and undistorted 3LSMO/3SIO supercells; (6) calculation details of the MAE of the distorted and undistorted 3LSMO/3SIO supercells; (7) layer- and element-projected MAE and charge distribution at interface calculations in the distorted and undistorted 3LSMO/3SIO supercells

## AUTHOR INFORMATION

### Corresponding Author

**Xianfeng Hao**-*Key Laboratory of Applied Chemistry, College of Environmental and Chemical Engineering, Yanshan University, Qinhuangdao 066004, China;* E-mail: xfhao@ysu.edu.cn

**Zhiming Wang**-*CAS Key Laboratory of Magnetic Materials and Devices, Ningbo Institute of Materials Technology and Engineering, Chinese Academy of Sciences, Ningbo 315201, China; Zhejiang Province Key Laboratory of Magnetic Materials and Application Technology, Ningbo Institute of Materials Technology and Engineering, Chinese Academy of Sciences, Ningbo 315201, China; Center of Materials Science and Optoelectronics Engineering, University of Chinese Academy of Sciences, Beijing 100049, China;* E-mail: zhiming.wang@nimte.ac.cn

### Author Contributions

Z. Lu conducted the data acquisition and drafted the manuscript, J. Liu participated in the sample fabrication and the data acquisition. L. Wen, P. Jiang, Z. Zhong and X. Hao performed the theory calculation. J. Feng, S. Kong, X. Zhen, S. Li, Z. Zhong, J. Zhu, X. Hao and Z. Wang contributed to the data interpretation. X. Hao, Z. Wang and R.-W. Li contributed to the data interpretation and manuscript writing.

### Notes

The authors declare no competing financial interest.

### ACKNOWLEDGEMENTS

This work was supported by the National Key Research and Development Program of China (Nos. 2017YFA0303600, 2019YFA0307800), the National Natural Science Foundation of China

(Nos. U1832102, 11874367, 51931011, 51902322), the Key Research Program of Frontier Sciences, Chinese Academy of Sciences (No. ZDBS-LY-SLH008), the Thousand Young Talents Program of China, the 3315 Program of Ningbo, the Natural Science Foundation of Zhejiang province of China (No. LR20A040001), the Public Welfare Technical Applied Research Project of Zhejiang Province (No. LY21E020007), the Junior Science Foundation of Zhejiang Province of China (No. LQ19A040002), the Ningbo Natural Science Foundation (No. 2019A610050, 2019A610055) , the Beijing National Laboratory for Condensed Matter Physics.

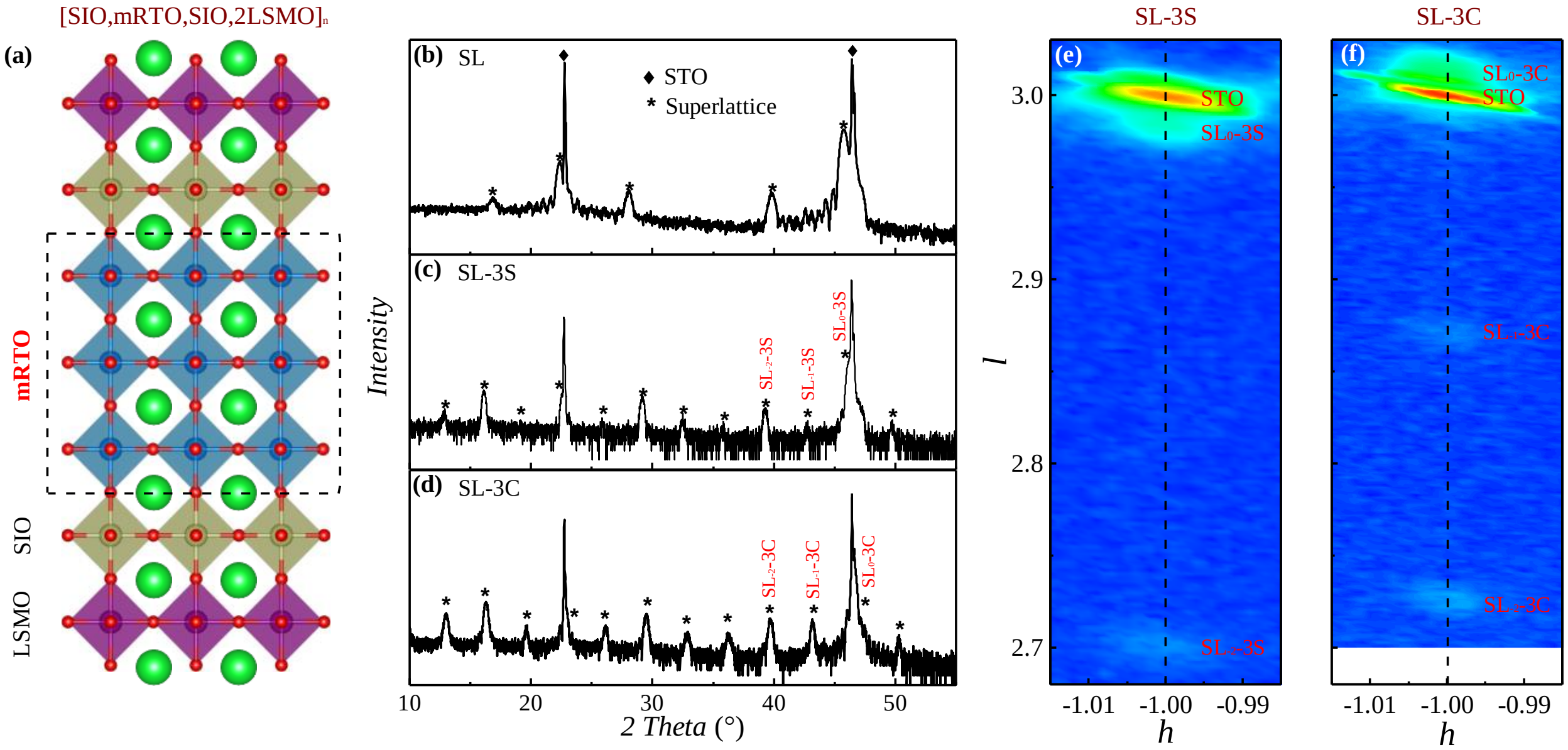

Figure 1. **Structure and epitaxial growth of [SIO,3RTO,SIO,2LSMO]$_n$ superlattice films,** where m represents the layer number of RTO in one building block and n indicates the repetition number of the building block. In one sense, every block is consisted of (SIO,mRTO,SIO) and 2-u.c. LSMO. (a) Schematic diagram of one building block. XRD diffractions of the films: (b) m=0, [2SIO,2LSMO]$_{10}$ (SL), (c) m=3, [SIO,3STO,SIO,2LSMO$_2$]$_{10}$ (SL-3S) and (d) m=3, [SIO,3CTO,SIO,2LSMO]$_{10}$ (SL-3C). Reciprocal space mappings of (e) [SIO,3CTO,SIO,2LSMO]$_{10}$ and (f) [SIO,3CTO,SIO,2LSMO]$_{10}$ film around (103) peak of the STO substrate. SIO, RTO and LSMO represent $SrIrO_3$, $SrTiO_3$ or $CaTiO_3$ and $La_{0.67}Sr_{0.33}MnO_3$, respectively. The green balls represent Sr, La and Ca atoms.

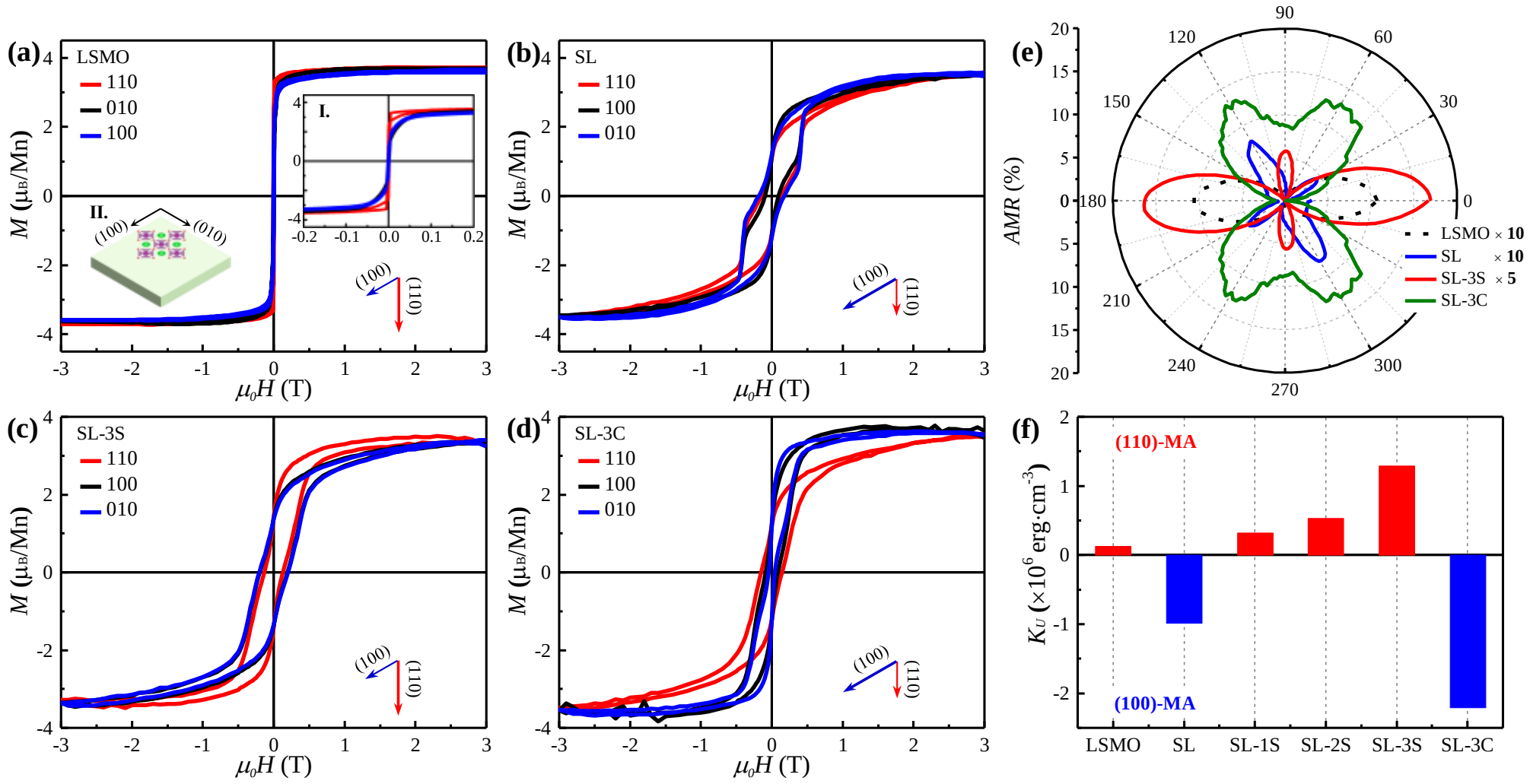

Figure 2. **In-plane MA of the LSMO and superlattices modulated by the (SIO,mRTO,SIO) building block**. In-plane magnetism loops along (100), (010) and (110) crystalline directions of (a) LSMO, (b) [2SIO,2LSMO]$_{10}$

(SL), (c) [SIO,3STO,SIO,2LSMO]$_{10}$ (SL-3S) and (d) [SIO,3CTO,SIO,2LSMO]$_{10}$ (SL-3C) superlattice films. The magnetic measurements are performed at 10 K. The inset I and II show the schematic of the measurement and magnified $\mu_0 H$-$M$ loop in (a), respectively. The blue and red arrows denote the magnetic axis along (100) and (110), respectively. (e) Polar plots of in-plane anisotropy magnetoresistance (AMR). The AMR values of the LSMO, SL and SL-3S films are magnified by 10, 10 and 5 times, respectively. The current and initial magnetic field are along the (100) direction and the magnetic field (3 T) is rotated within the film plane. (f) Dependence of magnetic anisotropy energy $K_U$ on the inserting STO and CTO layers and their thicknesses of m u.c. (m= 1, 2, 3). The positive sign of the $K_U$ corresponds to the (110) magnetic easy-axis while the negative one indicates a 45° rotation to (100).

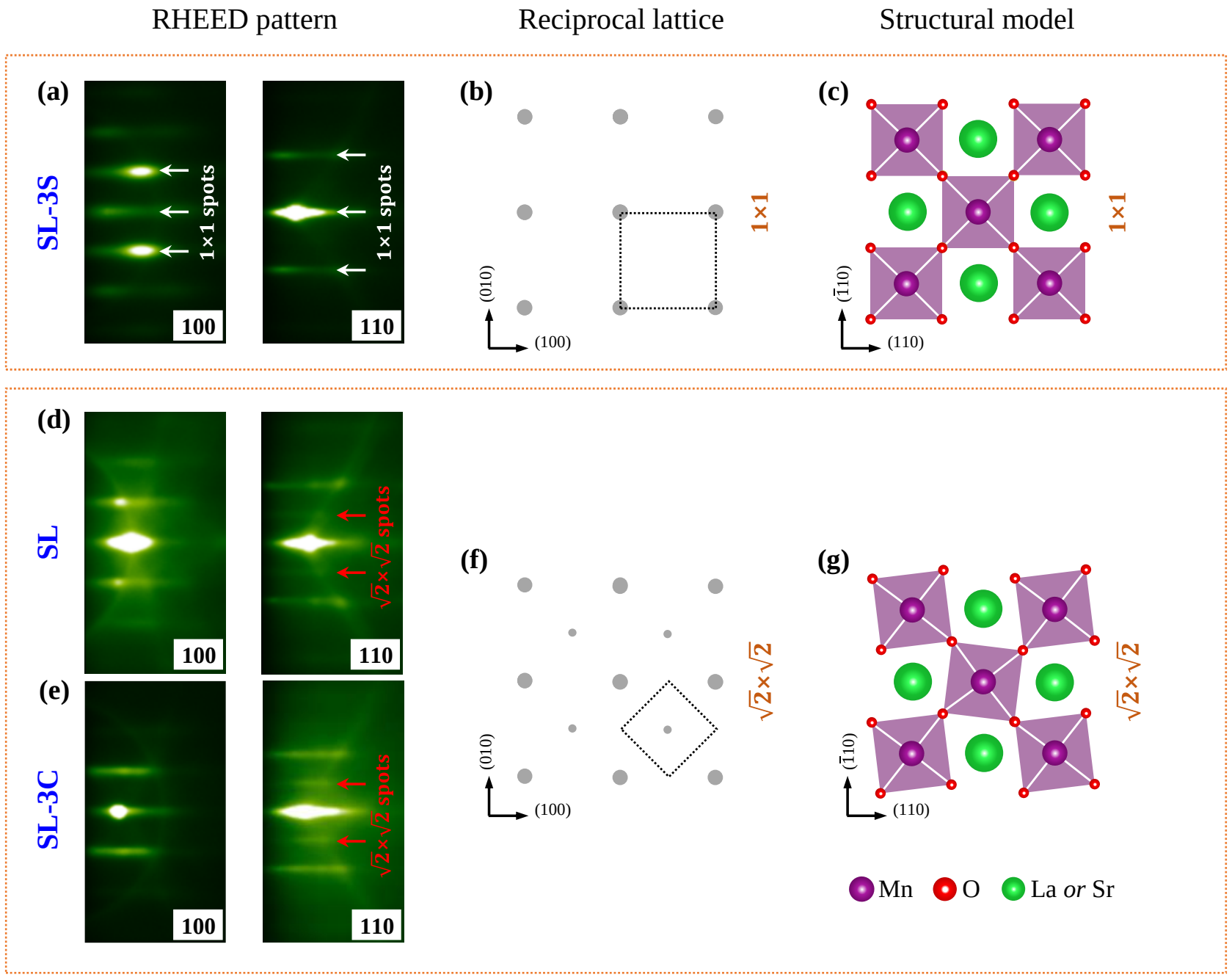

Figure 3. **In-plane structural transition of the [SIO,mRTO,SIO,2LSMO]$_{10}$ superlattice films modulated by STO and CTO**. The RHEED patterns of (a) [SIO,3STO,SIO,2LSMO]$_{10}$ (SL-3S), (d) [2SIO,2LSMO]$_{10}$ (SL) and (e) [SIO,3CTO,SIO,2LSMO]$_{10}$ (SL-3C) superlattice films after growth captured along (100) (left) and (110) (right) directions. The diffraction patterns marked with the red arrows in (d) and (e) exhibit a structural transition from a 1 × 1 (marked with white arrows) to a $\sqrt{2}\times\sqrt{2}$ (marked with red arrows) surface periodicity induced by altering the RTO. Reciprocal lattice patterns of (b) SL-3S and (f) SL and SL-3C films inferred from the RHEED patterns. (c) and (g) Structure models explaining the observed changes of the RHEED patterns dependent on the RTO, i.e. STO and CTO, respectively.

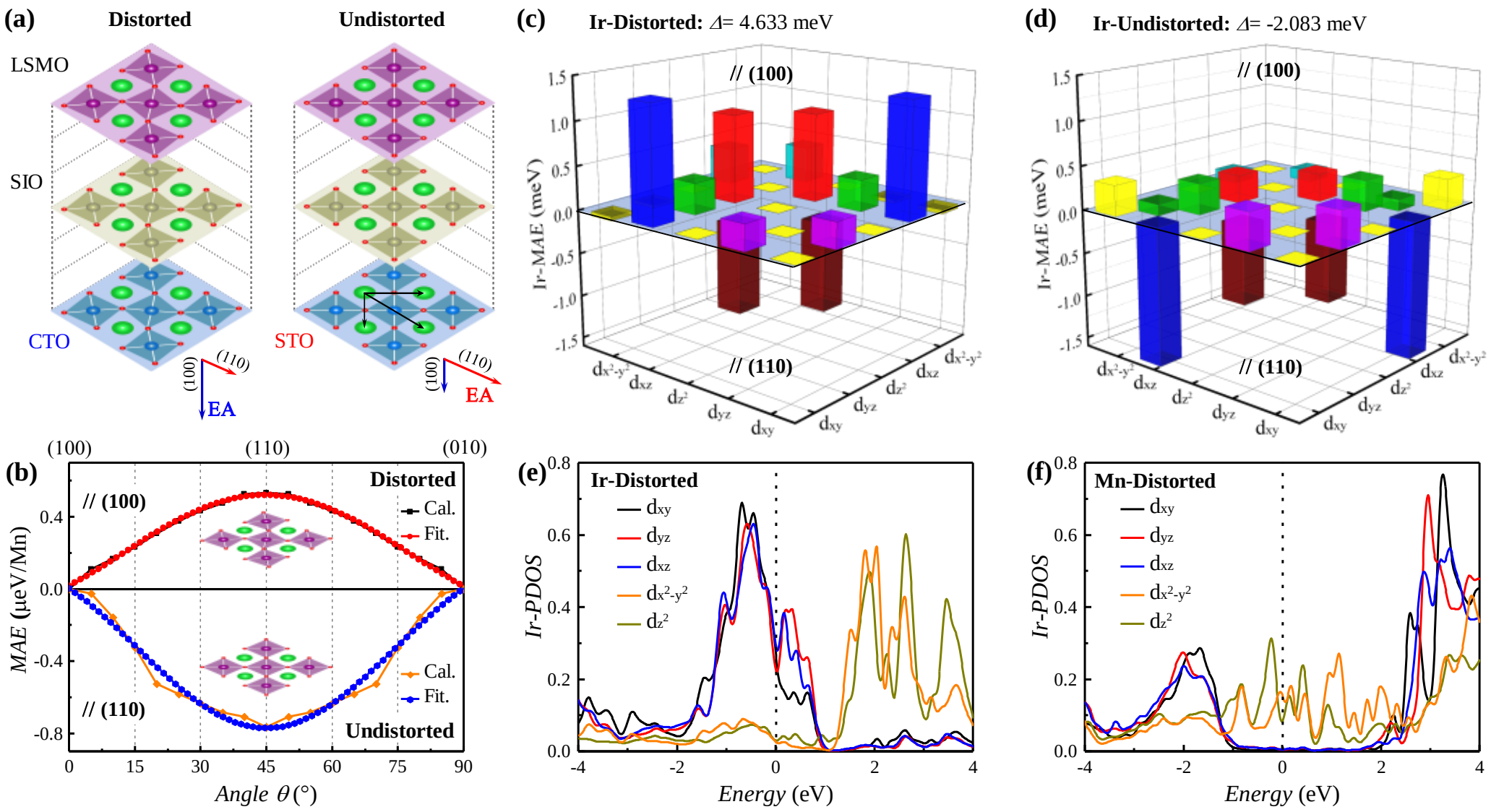

Figure 4. **Mechanism of the structure-dependent magnetic anisotropy (MA).** (a) The anisotropy of the superlattices with distorted (left) and undistorted (right) structures. (b) Simulation of the MAE for the distorted and undistorted 3LSMO/3SIO supercells by Bruno's Model. Cal. and Fit. are the abbreviations of calculation and fitting, respectively. The Angle $\theta$ denotes the angle between different crystal orientation and the (100)-axis in horizontal plane. The *d* orbit-resolved MAE of the Ir atoms in the (c) distorted and (d) undistorted supercells. The *d* orbit-resolved PDOSs of the (e) Ir and (f) Mn atoms in the distorted supercell. The symbol // presents the magnetic axis is parallel to the certain crystalline direction.

# Supporting Information

## I. Fabrication and MA of LSMO and $[2SIO,2LSMO]_{10}$ superlattice films

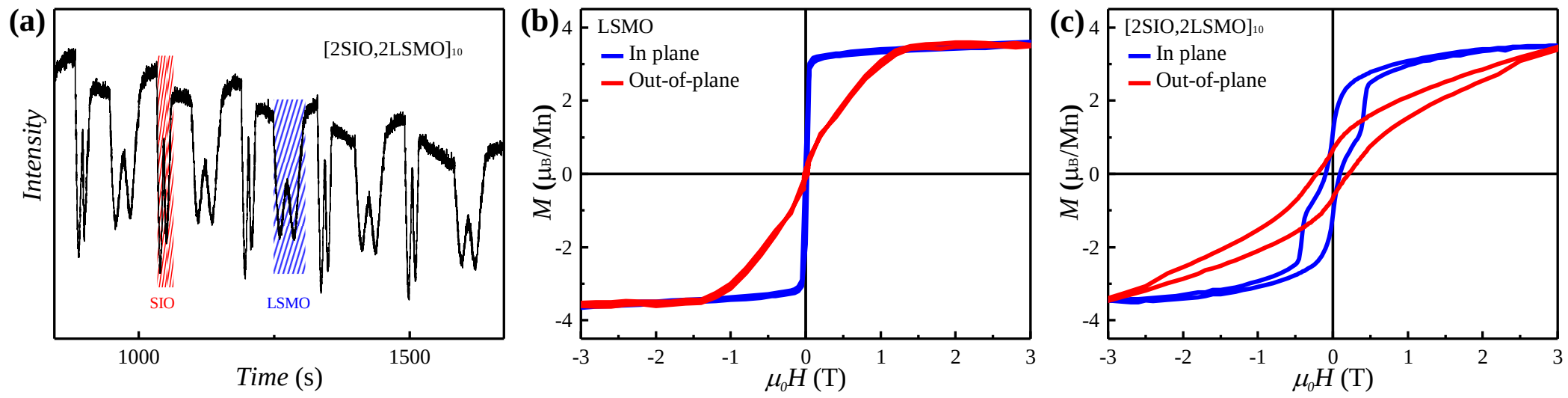

Figure S1. (a) The typical RHEED intensity oscillations of the LSMO and SIO during the deposition of $[2SIO,2LSMO]_{10}$ superlattice. In-plane and out-of-plane magnetism loops of (b) LSMO and (c) $[2SIO,2LSMO]_{10}$ superlattice films. The magnetic measurements are performed at 10 K.

## II. Structure and MA of $[SIO,STO,SIO,2LSMO]_{10}$ (SL-S) and $[SIO,2STO,SIO,2LSMO]_{10}$ (SL-2S) superlattice films.

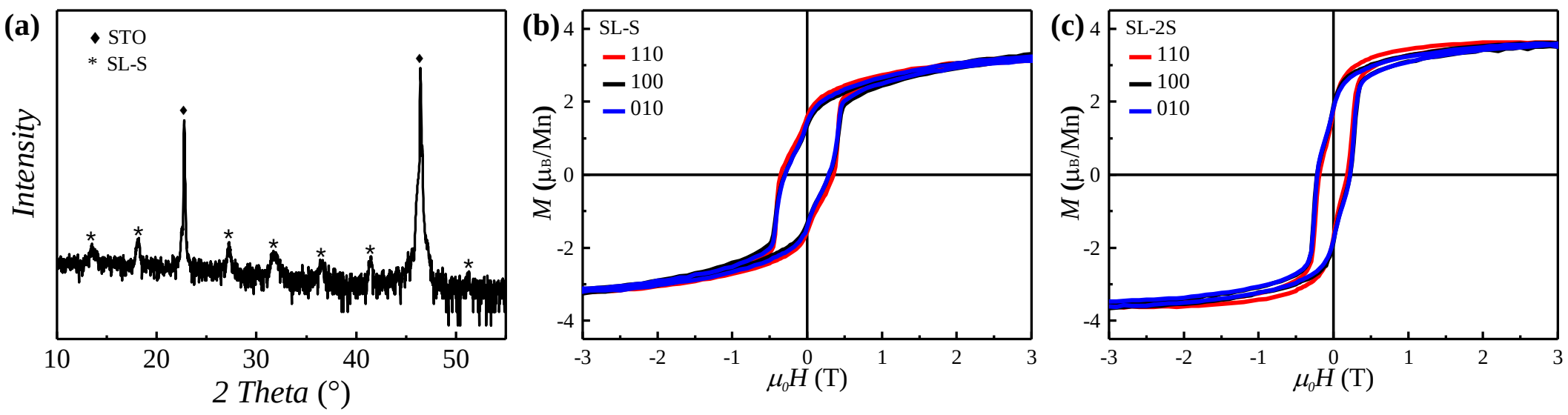

Figure S2. (a) XRD diffraction and (b) in-plane magnetic anisotropy of the SL-S, (c) in-plane magnetic anisotropy of SL-2S. The magnetic measurements are performed at 10 K.

## III. In-plane magnetic anisotropy of the LSMO and superlattice films

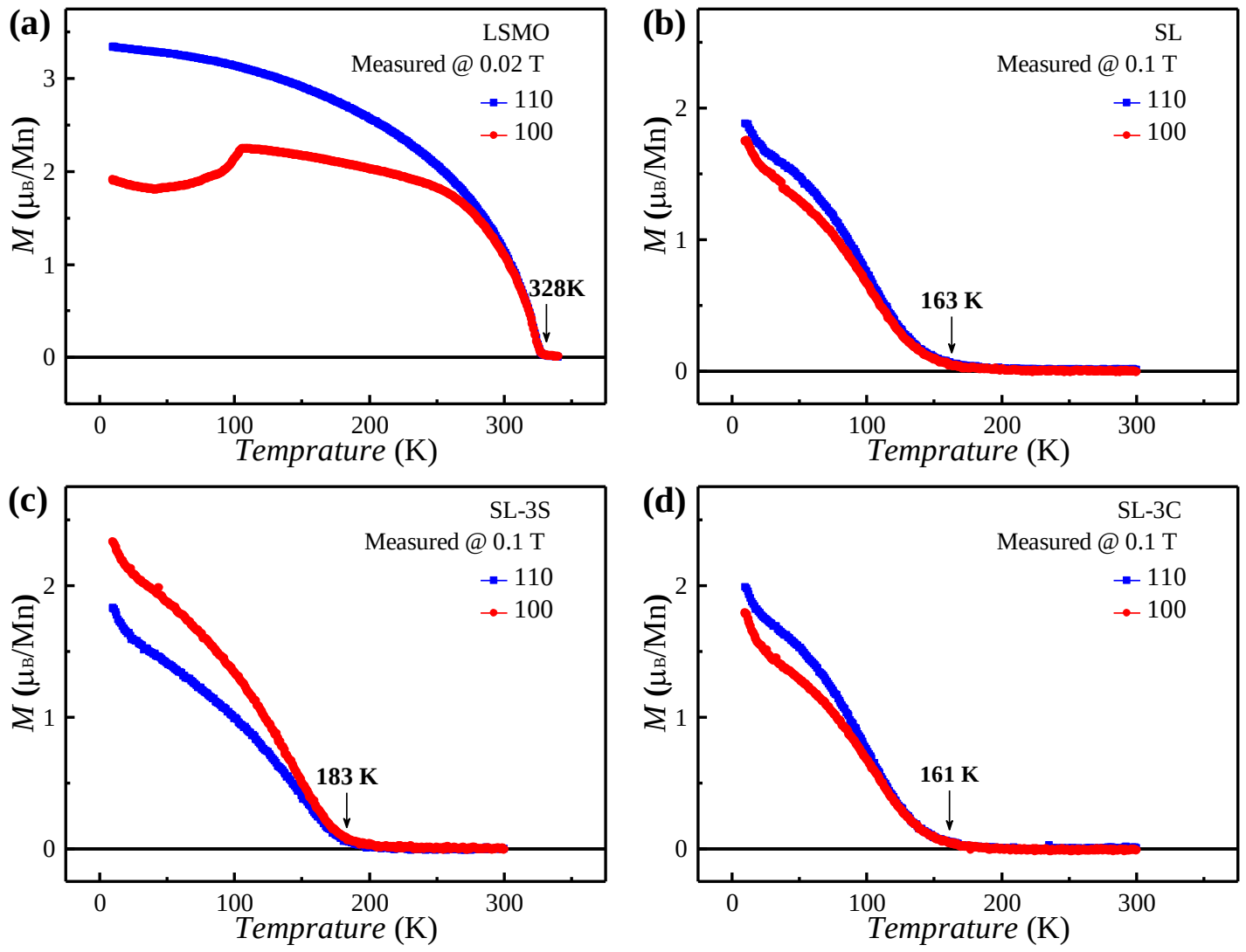

Figure S3. In-plane magnetization *v.s* temperature (*M-T*) curves along (100) and (110) crystalline directions of (a) LSMO, (b) $[2SIO,2LSMO]_{10}$ (SL), (c) $[SIO,3STO,SIO,2LSMO]_{10}$ (SL-3S) and (d) $[SIO,3CTO,SIO,2LSMO]_{10}$ (SL-3C) superlattice films. The measurements are performed at 0.02 T and 0.1 T after field cooling with 3T for the LSMO and superlattice films, respectively.

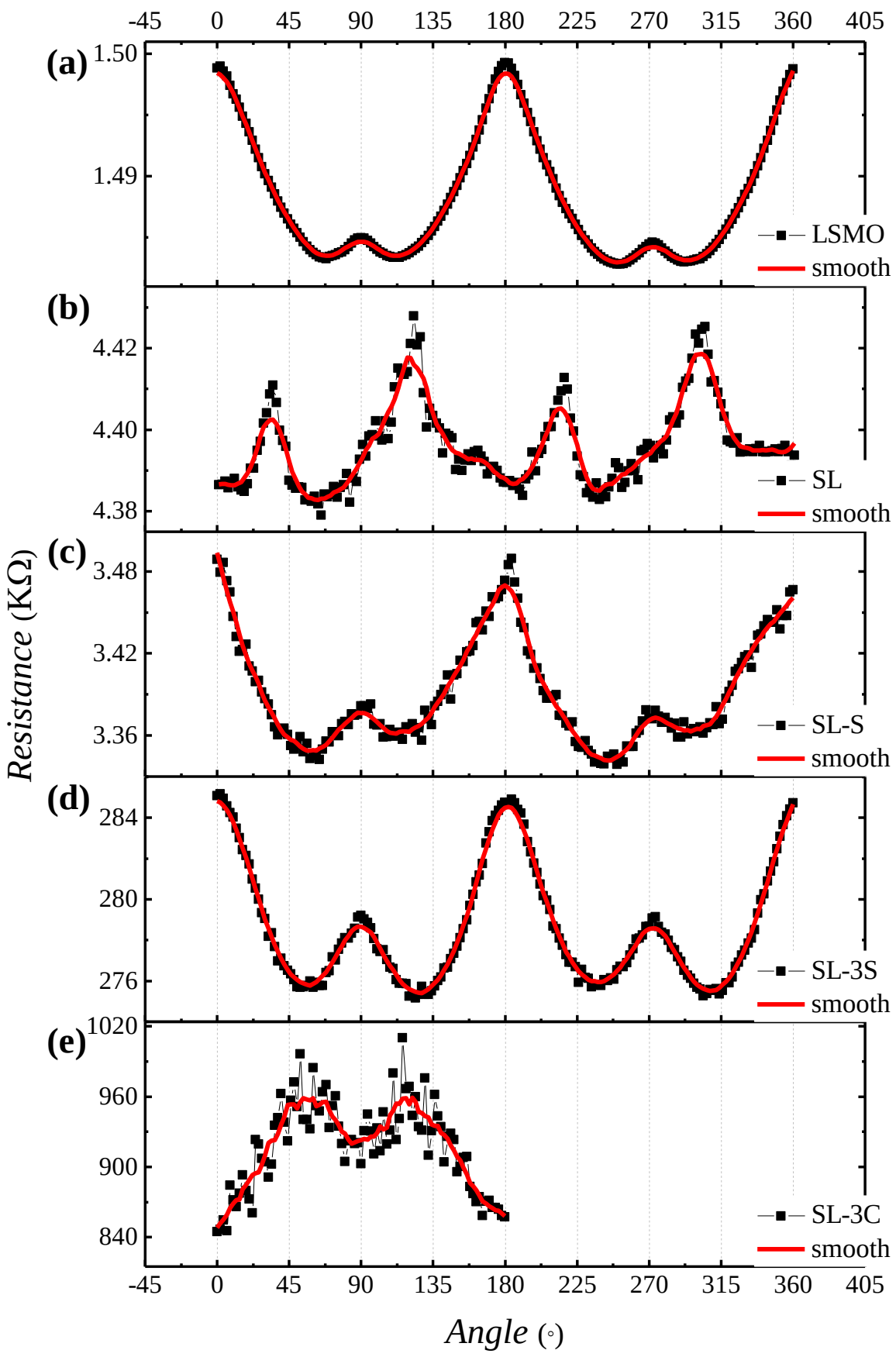

Figure S4. The in-plane anisotropy magnetoresistance (AMR) measured with in-plane magnetic field (3 T) rotating from 0° to 360° for (a) LSMO, (b) $[2SIO,2LSMO]_{10}$ (SL), (c) $[SIO,STO,SIO,2LSMO]_{10}$ (SL-S), (d) $[SIO,3STO,SIO,2LSMO]_{10}$ (SL-3S) and (e) $[SIO,3CTO,SIO,2LSMO]_{10}$ (SL-3C) superlattice films, respectively. The dot-line and solid line curves represent the experimental and smoothed results, respectively. The current and initial magnetic field are along the (100) direction and the magnetic field is rotated within the film plane.

## IV. In-plane structural transition of the SIO modulated by STO and CTO.

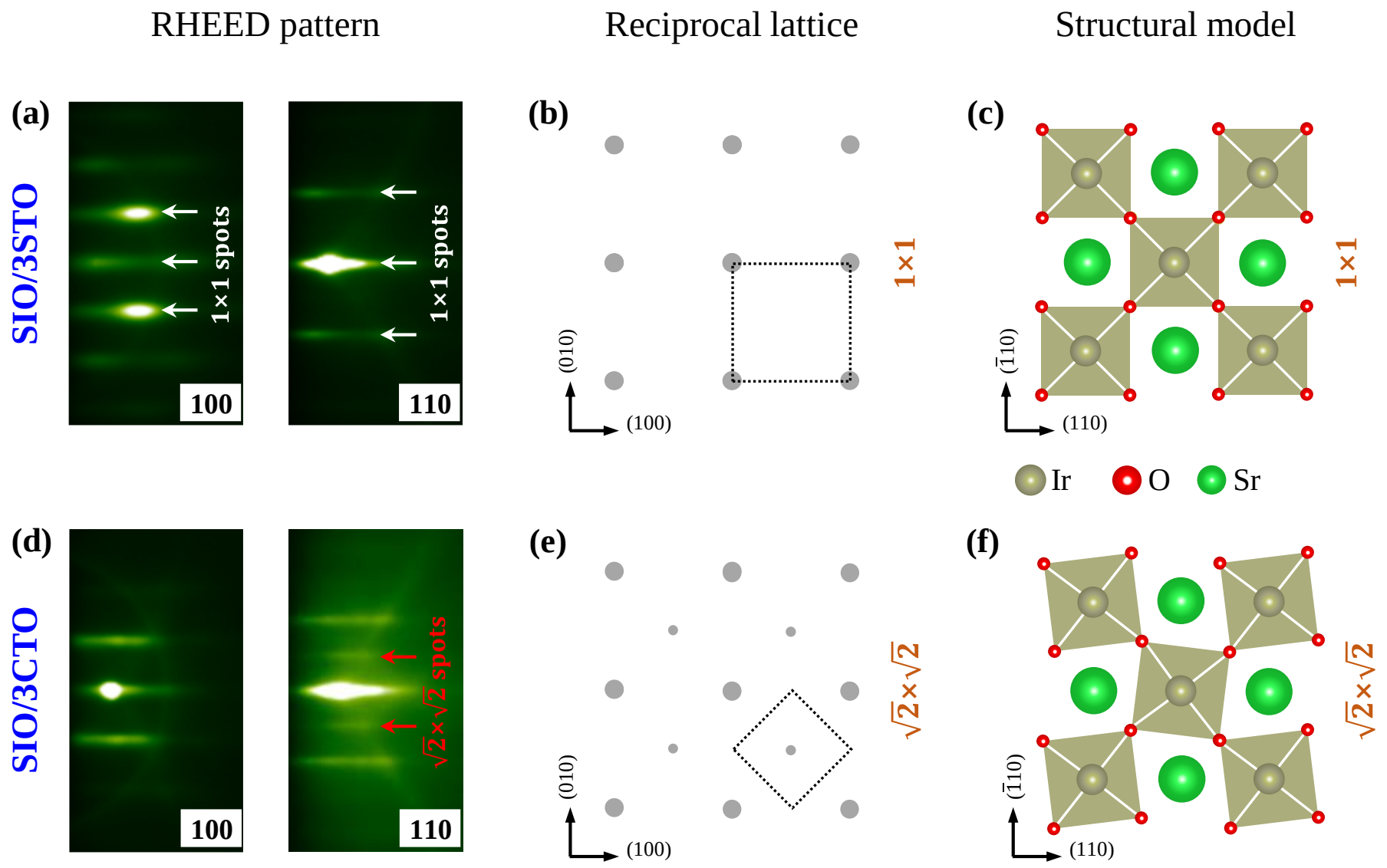

Figure S5. The RHHED patterns captured along (100) (left) and (110) (right) directions after growing (a) 3STO/SIO and (d) 3CTO/SIO bilayer films with SIO-terminated surfaces. The diffraction patterns marked with the red arrows in (d) and (e) exhibit a structural transition from a 1 × 1 (marked with white arrows) to a $\sqrt{2}\times\sqrt{2}$ (marked with red arrows) surface periodicity induced by altering the RTO. Reciprocal lattice patterns of (b) 3STO/SIO and (e) 3CTO/SIO films inferred from the RHEED patterns. (c) and (f) Structure models explaining the observed changes of the RHEED patterns dependent on the RTO, i.e. STO and CTO.

## V. Magnetic ground states for the distorted and undistorted 3LSMO/3SIO supercells.

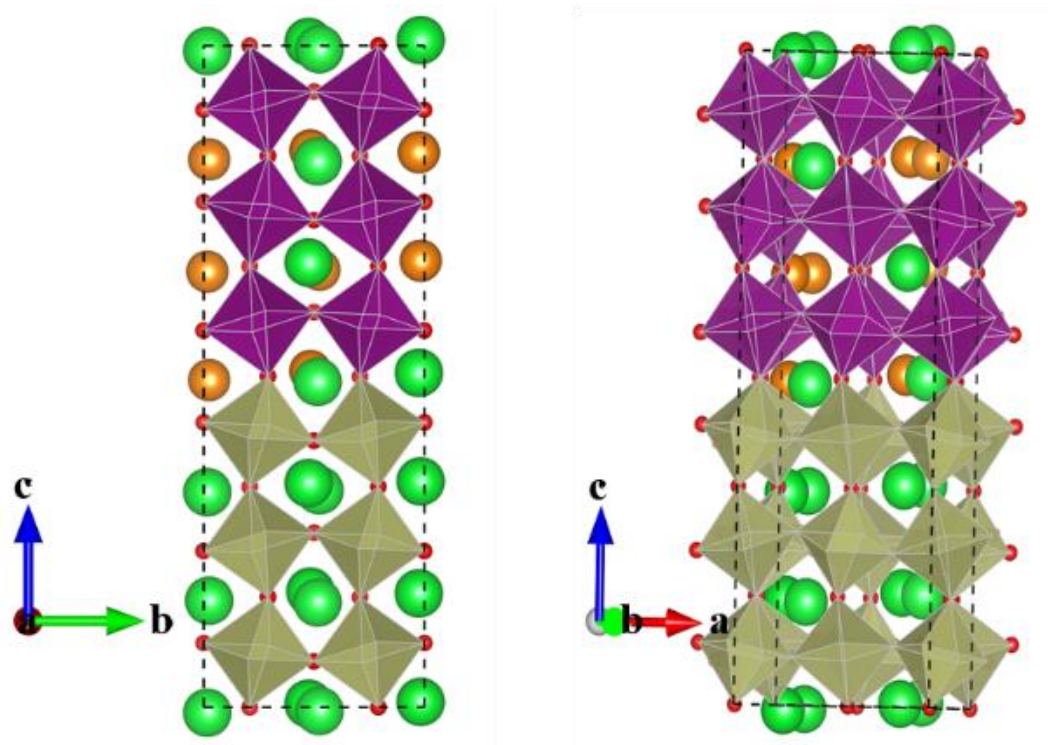

Figure S6. **Distribution of La and Sr cations in the LSMO layers of 3LSMO/3SIO supercells.** The gold and green spheres denote La and Sr atoms, respectively. The La and Sr distributions are random in the model, generated stochastically.

Figure S6 shows the distribution of La and Sr cations in the LSMO layers of the supercells. The gold and green spheres denote La and Sr atoms, respectively. Notice that, the La and Sr distributions are random in the model, generated stochastically. In experiments, the actual stoichiometry of the LSMO layers is $La_{0.67}Sr_{0.33}MnO_3$. Then, in order to capture this situation, we adopted the 3 layers of LSMO in the simulations, in which 8 out of 12 Sr atoms are randomly replaced by La atoms.

As is well known, $La_{1-x}Sr_xMnO_3$ exhibits the stability of ferromagnetic state due to the double exchange mechanism via degenerate $e_g$ orbitals[1], while the bulk $SrIrO_3$ is a semimetal with an extremely small number of charge carriers without any magnetic moment[2]. Concerning the magnetic properties of the interfaces, firstly the magnetic ground states for both two models were determined by comparing the total static energies within different magnetic configurations we considered in our simulations, i.e., the antiferromagnetic AFM-I one (FM LSMO layers and G-type AFM SIO layers), AFM-II one (AFM interactions between FM LSMO and FM SIO layers), ferromagnetic (FM-I) pattern (FM couplings between FM LSMO and FM SIO layers), as well as FM-II one (FM LSMO layers with non-magnetic configuration in SIO layers) as illustrated in Figure S7. The results are listed in Table 1. As is well known, $La_{1-x}Sr_xMnO_3$ exhibits the ferromagnetic state due to the double exchange mechanism via degenerate $e_g$ orbital, as evidenced by our temperature dependence magnetic susceptibility measurements illustrated in Figure S3. While the bulk $SrIrO_3$ is a semimetal with an extremely small number of charge carriers without any magnetic moment. In order to clarity the magnetic configuration within the SIO layers, we compared the total energies of four different magnetic alignments, provided that the LSMO layers are fixed to the ferromagnetic ordering. These four different configurations are the antiferromagnetic AFM-I one (FM LSMO layers and G-type AFM SIO layers), AFM-II one (AFM interactions between FM LSMO and FM SIO layers), ferromagnetic (FM-I) pattern (FM couplings between FM LSMO and FM SIO layers), as well as FM-II one (FM LSMO layers with non-magnetic configuration in SIO layers) as illustrated in Figure S7. The results are listed in Table 1. Notice that, the computed Ir magnetic moment is ~ 0.5 $\mu_B$, no matter what the magnetic configuration for SIO layers starts, suggesting the considerable magnetic interactions between the LSMO and SIO layers. For the distorted case, the most favorable magnetic state is AFM-II one, in which AFM interactions between FM LSMO and FM SIO layers. Exactly, the non-magnetic configuration convergences into the AFM-II case after the calculations. As for undistorted case, the ground magnetic state is FM case, however, the AFM-II case almost possesses the same total energy. Therefore, the magnetic anisotropy (MA) properties are performed within the AFM-II configuration.

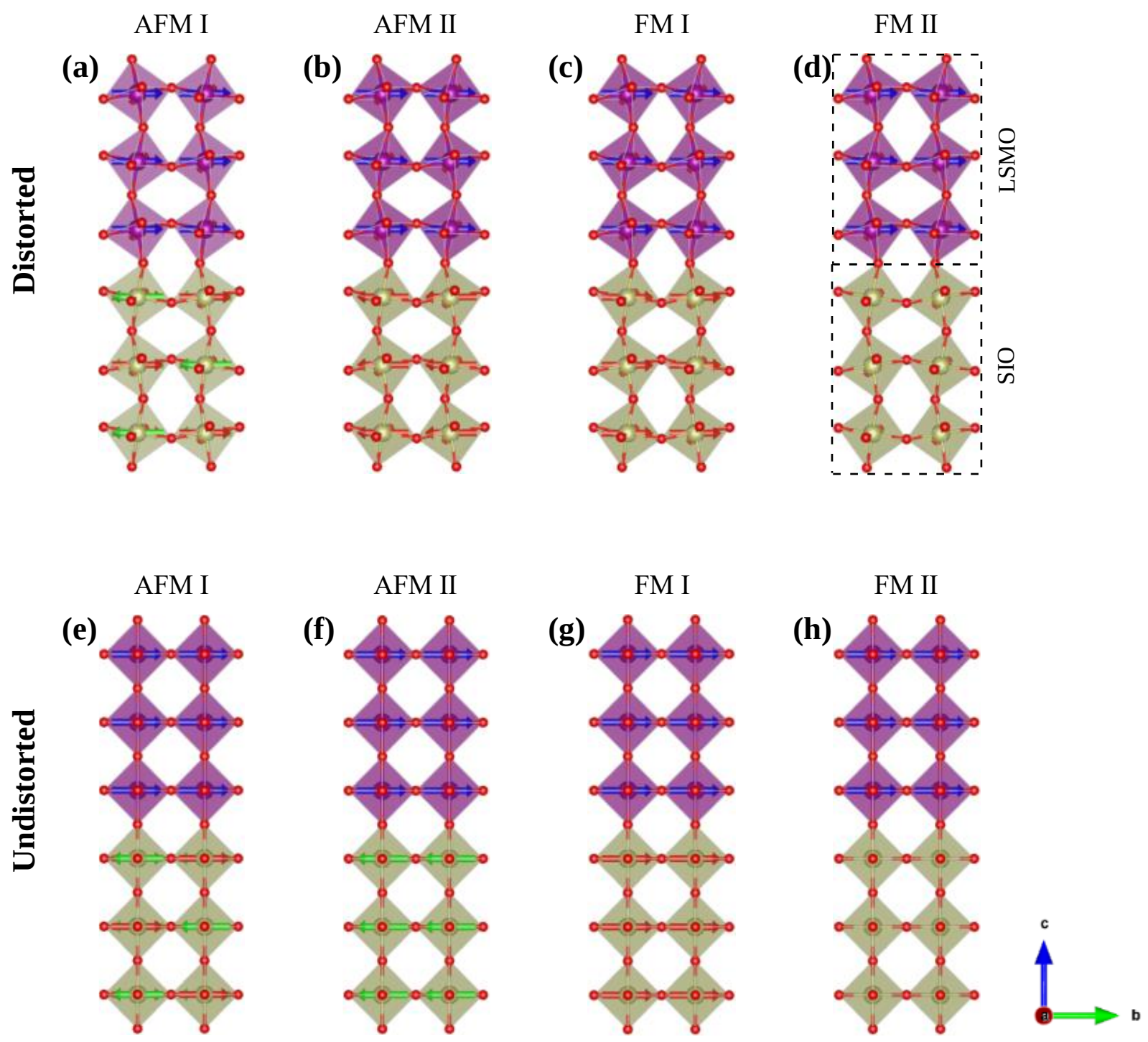

Figure S7. **Different magnetic configurations for the distorted and undistorted 3LSMO/3SIO supercells.** (a) and (e) antiferromagnetic (AFM-I) model (FM LSMO layers and G-type AFM SIO layers), (b) and (f) AFM-II model (AFM interactions between FM LSMO and FM SIO layers), (c) and (g) ferromagnetic (FM-I) model (FM couplings between FM LSMO and FM SIO layers), as well as (d) and (h) FM-II model (FM LSMO layers and non-magnetic SIO layers). The AFM and FM are the abbreviations of antiferromagnetic and ferromagnetic couplings, respectively. Distorted: (a-d), undistorted: (e-h).

Table 1. The total static energies within different magnetic configurations in the distorted and undistorted 3LSMO/3SIO supercells.

| **Magnetic structure** | **Distorted** | **Undistorted** |
|---|---|---|
| AFM I (SIO: G-AFM) | -835.2640 | -828.9141 |
| AFM II (SIO: FM) | -835.4841 | -829.0001 |
| FM I (SIO: FM) | -834.1993 | -829.0166 |
| FM II (SIO: PM) | -835.4840 | -828.9139 |

## VI. **MAE of the distorted and undistorted 3LSMO/3SIO supercells.**

Based on the magnetic configurations, we focused on the magnetic anisotropy energies of rotational and irrotational structures using the force theorem, which consists of three consecutive steps. First, the self-consistent calculations without SOC coupling are conducted to obtain the charge density of the ground state. And then, the charge density achieved before is adopted as an input to carry out the non self-consistent calculation with SOC effect for obtaining the total energy of the system for specific magnetization directions, such as (100), (010) and (110). Having established the magnetic easy-axis, the MA properties of LSMO-SIO layers is evaluated by the variance of the static energies as a function of the angle $\theta$, which is the rotation angle between different magnetization directions. The results are illustrated in Figure 4b. Evidently, the easy magnetization axis for rotational 3LSMO/3SIO case is (100), while the corresponding one for irrotational case is (110), which are in excellent agreement with the experimental observations. Furthermore, the angular dependence of the magnetic energies for both cases is well fitted to the conventional uniaxial anisotropy Bruno's model. Notice that, in order to directly compare to the experimental data, the theoretical MAE ones are the average values of crude static energies of the individual $\theta$ and its corresponding complement angle (90-$\theta$), due to the fourfold rotation symmetry magnetic domains induced by the fourfold symmetry structure[3, 4].

VII. **Layer- and element-projected MAE and charge distribution at interface calculations in the distorted and undistorted 3LSMO/3SIO supercells**

We perform layer-, element- and orbital-projected MAE calculations of the 3LSMO/3SIO supercell based on a second-order perturbation theory. As shown in Figure S8, the layer-projected MAE calculations of the SIO show that the top (bottom) interfacial SIO has an energy of 4.306 (4.960) meV multiplied almost three times than the SIO interlayer. It indicates that the MAE of the SIO is mainly contributed by the interfacial SIO. Figure S9a-d show wholly the element-projected MAE *d* orbital matrix elements for the distorted and undistorted structures. Apparently, the predominant alternation of Ir-element orbital matrix is originated from the matrix elements ($d_{xy}$, $d_{xz}$), i.e., positive in distorted case while negative in undistorted one, giving rise to the switch of the single ion anisotropy of the SIO. This transformation can be traced back to uneven occupation of the $d_{xz}$ and $d_{yz}$ orbitals in the distorted structure (density of state plots for Ir atom in Figure 4e), which is also evidenced from the differential charge density plots as presented in Fig. S10, where the asymmetrical charge distribution is clearly visualized, compared to the undistorted one. On the other hand, although the sign of the MAE value of Mn cations remains unchanged, the change mode of Mn-element orbital matrix are closely associated with the pattern of Ir-element, suggesting the covalent interaction of Mn-O-Ir bonds across the interface. In addition, the density of states illustrated in Fig. 4e-f and Fig. S9e-f reveal the hybridization between Mn and Ir cations across the interface in terms of the similar curve shape and energy range of $d_{yz}$, $d_{xz}$ and $d_{z2}$ orbitals. Furthermore, this covalent interaction is also uncovered from the differential charge density plots, in which the significant charge transfer or distribution are concentrated on the interface section.

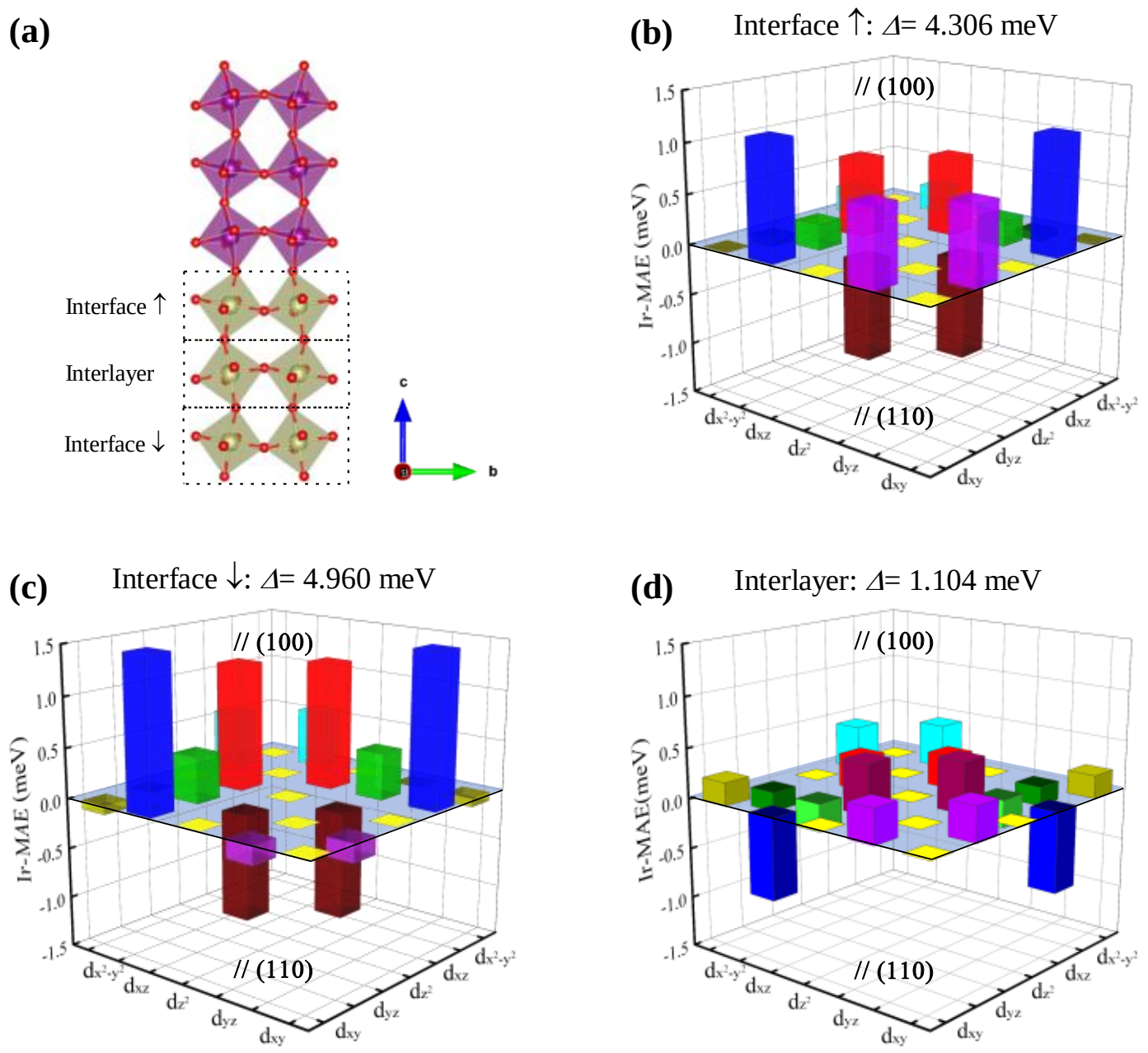

Figure S8. **Layer-resolved MAE of the Ir atoms in the distorted and undistorted 3LSMO/3SIO supercells.** (a) Structural configuration. The MAE of the SIO at (b) top and (c) bottom interfaces, and (d) the SIO interlayer. The symbol // presents the magnetic axis is parallel to the certain crystalline direction. The symbol $\Delta$ is the magnetic anisotropy energy (MAE).

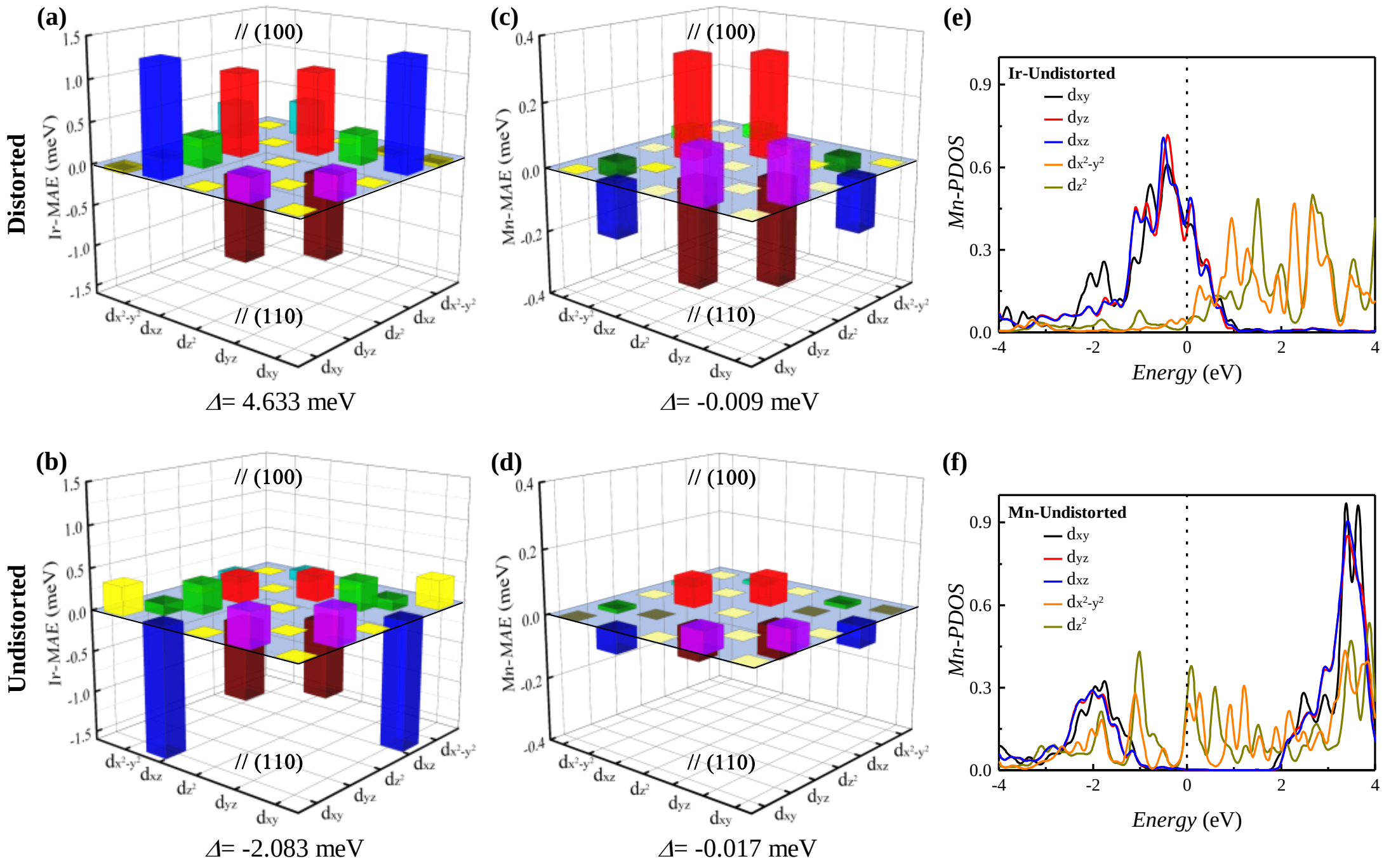

Figure S9. **Element-resolved MAE and PDOS of the Ir and Mn atoms in the distorted and undistorted 3LSMO/3SIO supercells.** The MAEs of the atoms in the distorted (a) SIO and (c) LSMO, of the undistorted (b) SIO and (d) LSMO. The PDOS of the (e) Ir and (f) Mn atoms in the distorted supercell.

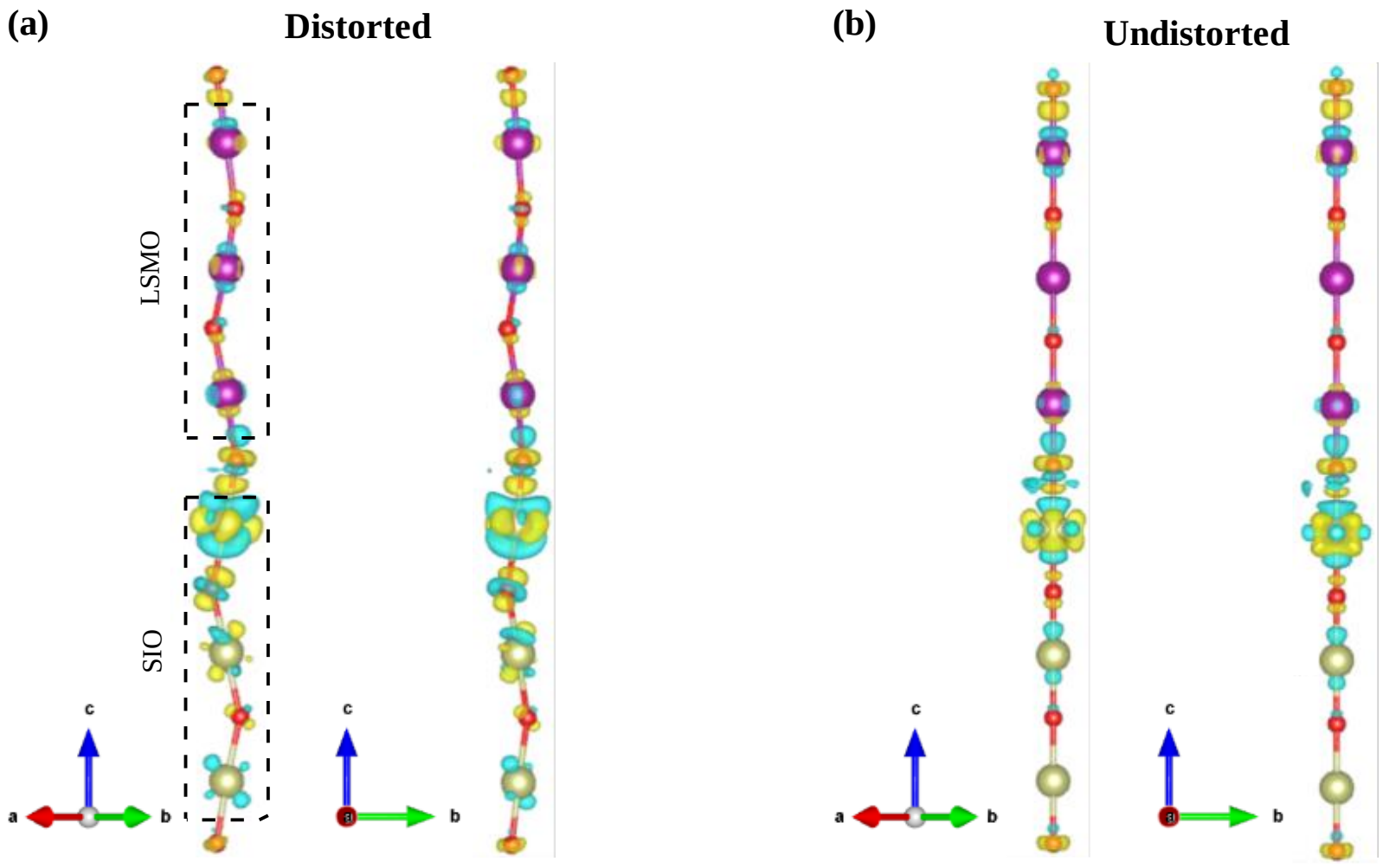

Figure S10. **The total charge density difference among interfacial Mn (purple), O (red), and Ir (dark yellow) atoms:** in the (a) distorted and (b) undistorted 3LSMO/3SIO supercells viewed from (110) (left) and (100) (right) directions. The yellow and wathet blue distributions denote the electron accumulation and depletion regions, respectively.